# Affordable and Energy-Efficient Cloud Computing Clusters: The Bolzano Raspberry Pi Cloud Cluster Experiment

Pekka Abrahamsson, Sven Helmer, Nattakarn Phaphoom, Lorenzo Nicolodi, Nick Preda, Lorenzo Miori, Matteo Angriman, Juha Rikkilä, Xiaofeng Wang, Karim Hamily, Sara Bugoloni

Free University of Bozen-Bolzano

Piazza Domenicani 3, 39100 Bolzano, Italy

{firstname.surname}@unibz.it

*Abstract*—We present our ongoing work building a Raspberry Pi cluster consisting of 300 nodes. The unique characteristics of this single board computer pose several challenges, but also offer a number of interesting opportunities. On the one hand, a single Raspberry Pi can be purchased cheaply and has a low power consumption, which makes it possible to create an affordable and energy-efficient cluster. On the other hand, it lacks in computing power, which makes it difficult to run computationally intensive software on it. Nevertheless, by combining a large number of Raspberries into a cluster, this drawback can be (partially) offset. Here we report on the first important steps of creating our cluster: how to set up and configure the hardware and the system software, and how to monitor and maintain the system. We also discuss potential use cases for our cluster, the two most important being an inexpensive and green testbed for cloud computing research and a robust and mobile data center for operating in adverse environments.

## I. INTRODUCTION

Cloud services have become commonplace in day-to-day life, which is reflected by the ever-increasing use of social media applications and services for e-mail, storage, and video streaming. Facebook has claimed to have passed the 1 billion mark for active users, Apple has confirmed 50 billion downloads in their App Store, and almost half a billion users have a Gmail account. Companies are also transferring their services into the cloud, despite the claimed risks associated with this, such as privacy and ownership disputes. This is especially true for startups, which avoid creating a traditional computing infrastructure and rely on highly scalable and highly available cloud services instead. One could even claim that the Cloud slowly seems to be becoming a synonym for the Internet.

Nevertheless, there are still numerous issues in the field of cloud computing. One important one that is of particular concern to us is the cost of setting up and operating large data centers. This includes the cost of the hardware, maintenance, and the electricity to run the computers and to cool them. A current challenge faced by High-Performance-Computing (HPC) Clouds, is how to build a system delivering an exaflop of computing power consuming only 20 MW of power [1]. Affordability and power consumption are also critical when it comes to deploying systems in developing countries, which are lacking the financial means and the infrastructure to deploy large and costly data centers. Unless this is addressed, the global digital divide will increase even further. Another issue we are interested in is the mobility of computing clusters. Getting a huge amount of data to a data center can be a problem, due to bottlenecks in the communication infrastructure. In some cases it may be more cost-efficient to move a computing center to the data as illustrated by the ICE Cube modular data centers of Silicon Graphics [2]. In a similar way, it may make sense to move data centers close to energy sources for energy efficiency reasons. Mobility is also an important aspect when it comes to deploying data centers in disaster areas in case of emergencies or when moving systems to remote areas in undeveloped regions.

These challenges provide great opportunities for researchers in academia and industry. Indeed, governments and companies world-wide are investing heavily in this research area. Unfortunately, today's students, who are the researchers of tomorrow, are not getting enough exposure to real cloud computing infrastructure. It is not enough to catch a glimpse of high-performance computing equipment while visiting a data center or to interact on a very abstract level with the Cloud. Students need hands-on experience in this area, which means that universities need to provide access to a suitable cloud computing infrastructure that can be used for experimentation, research, and teaching. Building and running this kind of infrastructure is a very costly endeavor for a university. For example, the University of Helsinki recently acquired a cloud computing cluster called Ukko (tinyurl.com/ukko1) with 240 Dell PowerEdge M610 nodes, each with 32GB of RAM and 2 Intel Xeon E5540 2.53GHz quadcore CPUs. Based on publicly available information, the cost of the hardware was close to one million Euros plus another substantial amount of money for the cooling system. Additionally, changes to the structure of the building housing the cluster had to be made. On top of this we have the energy bills: in the case of the Ukko computing cluster the power consumed when processing heavy workloads reached 115 KW for the whole data center and 68 KW for the cluster itself (tinyurl.com/ukko3). Compared to commercial data centers, which can easily cost up to a hundred million Euros, it may seem a small sum (and it is great if an organization is able to afford building a real cloud data center), but for many universities and schools this kind of investment is still out of reach.

The question we set out to answer was: how can we develop an affordable, energy-efficient and portable cloud-computing cluster? In the ideal case we would like to have a cluster with a low or even zero carbon footprint, individual elements that are cheap to replace (costing only tens of dollars each), and a portable system that can be easily taken apart and re-assembled. Clearly, we have not yet come up with a final answer to our question, but we have made an important first step by building a cluster composed of 300 Raspberry Pi computers, which, at the time of writing, was the largest of its kind. The Raspberry Pi is ideal for our purposes, due to its very reasonable price, low energy consumption, and small size. At the same time it still provides decent performance, especially when looking at the computing power offered per watt.

During the construction of our cluster, students from the Faculty of Computer Science at the Free University of Bozen-Bolzano configured the hardware and software to get the system up and running, while students from the Faculty of Design and Art worked on the design of the racks housing the individual Raspberry Pis, the switches, and the power supplies. While this was and is already a great learning experience for everyone involved, we expect many more people to participate in using the system as a low-cost and green testbed for cloud computing research.

The remainder of the paper is organized as follows. In the next section, we summarize the basic properties of Raspberry Pi computers and also discuss their known technical limitations. Section 3 outlines current RPi cluster projects. The objective of the RPi cloud cluster developed at the Free University of Bolzano is presented in Section 4. This is followed by a description of the hardware configuration and the system software of the cluster. We conclude with a summary of our current achievement and directions for future work.

## II. THE RASPBERRY PI COMPUTER

The Raspberry Pi Foundation developed a credit-card-sized, single-board computer called Raspberry Pi (RPi), with the aim of stimulating the teaching of computer science at school level. The project was initiated by Eben Upton, a British engineer, in 2006 to tackle the lack of programmable hardware for children and in 2009 the Raspberry Pi Foundation, a charitable organization, was officially founded [3]. In less than two years an alpha version of the Raspberry Pi, used primarily as a development platform and for demonstration purposes, was released. This was followed by a beta version in December 2011. The Foundation was caught off guard by the huge demand when online sales started on 29 February 2012: it was reported that customers needed to wait for hours to submit a preorder.

### A. System Specification

The heart of an RPi is an integrated circuit (system on chip) combining an ARM 700 MHz processor (CPU), a Broadcom VideoCore graphics processor (GPU), and 256/512 MB of RAM on a single chip [4]. Moreover, there is an SD card slot for storage and I/O units such as USB, Ethernet, audio, RCA video, and HDMI. Power is provided via a micro-USB connector. The RPi comes in two variants, A and B, with the model B offering features that the model A lacks: it has 512 MB of RAM (instead of 256 MB) and an onboard Ethernet port. Table I summarizes the hardware specification of model B and Figure 1 gives a schematic overview of the system.

| System on chip | Broadcom BCM2835 |
|---|---|
| CPU | 700 MHz Low Power ARM1176JZ-F Processor |
| GPU | Dual Core VideoCore IV Multimedia Co-Processor |
| Memory | 512MB SDRAM |
| Ethernet | Onboard 10/100 Ethernet RJ45 jack |
| USB | Dual USB 2.0 Connector |
| Video output | HDMI (rev 1.3 & 1.4) Composite RCA (PAL and NTSC) |
| Audio output | 3.5mm jack, HDMI |
| Storage | SD, MMC, SDIO card slot |

TABLE I. HARDWARE SPECIFICATION (MODEL B)

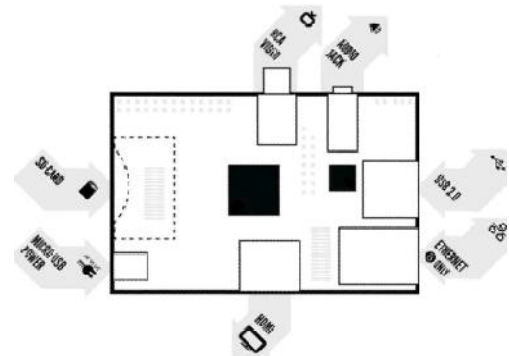

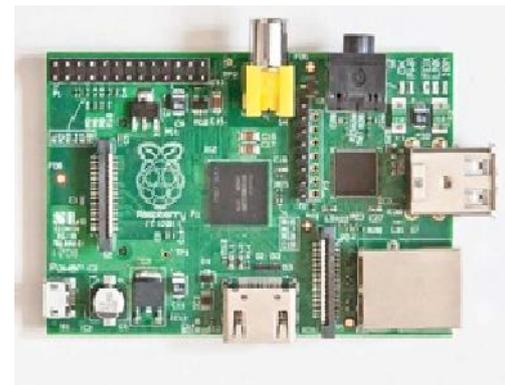

Fig. 1. Schematic overview

In terms of system software, the Raspberry Foundation recommends using Raspbian as operating system (OS). Raspbian is a port of the well-known Linux distribution Debian optimized for the ARMv6 instruction set, e.g. providing better performance for floating point arithmetic operations [5]. Currently, Raspbian comes with around 35,000 prebuilt packages, simplifying the process of installing software on an RPi. Other operating systems are (or will soon be) available for RPi as well, this includes Android, Arch Linux ARM, Firebox OS,

| Device | time to complete |
|---|---|
| RPi 700 MHz | 107 ms |
| RPi 800 MHz | 93 ms |
| RPi 900 MHz | 82 ms |
| SheevaPlug | 78 ms |
| E8400 | 0.85 ms |

TABLE II. CPU SPEED

| Device | pages/sec | power in watts |
|---|---|---|
| RPi | 17 | 3 W |
| Kirkwood | 25 | ≈13 W |
| MK802 | 39 | ≈4 W |
| Atom 330 | 174 | 35 W |
| G620 | 805 | 45 W |

TABLE III. HTTP SERVER BENCHMARK

Google Chromium, Fedora, Plan 9 from Bell lab, RISC OS, and UNIX [4].

*B. Limitations*

The small size and low power consumption come at a price, though. In terms of performance there are limitations to what a single RPi can accomplish. The Raspberry Foundation provides some comparisons between RPis and other devices. Its overall performance is roughly comparable to a PC using an Intel Pentium 2 processor clocked at 300MHz, and the graphics capabilities are roughly the same as the original Microsoft Xbox games console. In the following we are highlighting some areas that have been investigated in the context of performance.

*1) Processor Speed:* The RPi has been criticized for its outdated ARM11 processor running at 700 MHz and the separation of the L2 cache, which is reserved for the GPU. In April 2012, the computing performance of an RPi running Debian OS was tested using the SysBench benchmark suit [6]. The test calculated prime numbers up to a selected maximum based on 64-bit integer calculations. The test compared the RPi (model B) 700 MHz, an RPi (model B) overclocked to 800 and 900 MHz, a Marvell SheevaPlug (Ferroceon ARM chip running at 1.2GHz), and an average PC (Intel Core 2 Duo E8400 at 3GHz). The results are shown in Table II. While it is possible to "tune" an RPi by overclocking the CPU from 700 MHz to 900 MHz and the RAM clock from 400 MHz to 440 MHz, this will result in a shorter lifespan.

*2) SD Card Performance:* The SD card performance of an RPi is rather slow [7]. This can be partly explained by the general design of solid state flash memory. Read/write operations typically access memory in blocks (for example of 128 KB each). Even changing a single byte means reading a whole block, changing the byte, and rewriting a new block (which has to be prepared for writing). Additionally, the lifespan of an SD card is reduced significantly in the case of very frequent write operations. Consequently, some users connect external hard drives to their RPis. However, this increases the power consumption of the overall system.

*3) Performance in General Use Cases:* A test assessing the performance of RPis for general purpose computing was performed by bit-tech.net [6]. In this test, an RPi runs a lightweight Linux desktop environment (LXDE) on Debian OS. It took thirteen seconds for the OS to boot, eight seconds to load the desktop, six seconds to load the Midori web browser, and eight seconds to load the bit-tech homepage.[1] Scrolling on complex pages had a latency of several seconds, while loading Gimp (the GNU image manipulation program) took 87 seconds.

In another experiment the suitability of an RPi for serving HTTP pages was investigated [8]. The test web page represented a simple blog without database access (the postings were stored in a file directory). The total size of the page was 64.9 KB. An RPi (model B) was compared to a 1.2 GHz Marvell Kirkwood, a 1 GHz MK802, a 1.6 GHz Intel Atom 330, and a 2.6 GHz dual core G620 Pentium. All systems had a wired 1 GB Ethernet connection (which the Raspberry, having a 10/100 MBit ethernet card, could not utilize fully) and ApachBench[2] was used to execute the benchmark. The test consisted of requesting the page 1000 times (10 of the requests were run concurrently). Table III illustrates the result, in terms of speed and power consumption.

*4) Summary:* It is possible to use a (single) RPi to handle a light workload. It comes at a very affordable price and with a low power consumption. Clearly, a downside is the performance, in the following we discuss how to overcome this drawback by combining multiple RPis into a cluster.

III. CURRENT RASPBERRY CLUSTER PROJECTS

We are not the first to build an RPi cluster, there are a couple of projects that have put together RPis to form a larger entity. These projects, which we describe briefly in the following, however, are all on a smaller scale.

*A. Southampton Supercomputer*

A team led by Prof Simon Cox at the University of Southampton started a project to encourage students to work in the area of high performance computing. Details on the project can be found at http://www.southampton.ac.uk/~sjc/raspberrypi/pi_supercomputer_southampton.htm.

On the hardware side, 64 RPis are linked together to form a cluster with a total of 64 processors, 1TB of memory at a cost of under £2500. Each RPi is powered through a mobile charger and is connected to a switch via its Ethernet port. The racks to house the RPis were built using LEGO blocks. In terms of software, the RPis run Raspbian and on top of this MPI (Message Passing Interface) for communicating over the Ethernet connection. This allows the nodes to coordinate their actions and to distribute the workload.

*B. Boise State Beowulf Cluster*

Joshua Kiepert, a PhD student from Boise State University, built his own Beowulf cluster using RPis with the help of fellow students. The motivation for this was to have unlimited access to a cluster needed for doing computations in the context of his thesis. Details on this project are available at http://coen.boisestate.edu/ece/raspberry-pi/.

---

[1] http://www.bit-tech.net/

[2] http://httpd.apache.org/docs/2.2/programs/ab.html

The cluster consists of 33 RPis, 32 of which are doing the actual data processing, while one of them is used as a master. All of them are connected to a common switch and in addition to this, the master RPi also has access to NFS file systems. The cluster is running Arch Linux ARM with MPI on top of it.

*C. Glasgow Cloud Data Center*

This project was initiated at the School of Computer Science, University of Glasgow, with the purpose of constructing a warehouse-sized data center on a small scale with RPis [9]. This comes closest to what we have in mind, since it is about providing a research and education environment for gaining experience with the infrastructure and the management of a cloud data center.

At the moment, the hardware consists of 56 RPis housed in four LEGO mini-racks. Each rack is connected to a sixteen-way switch: fourteen sockets are used to hook up the RPis and two for inter-rack communication. On the software side, the RPis are running Raspbian, each node hosting three LXCs (LXC stands for Linux Container, which is a light-weight OS-level virtualization). According to the team's blog, they are currently experimenting with libvirt and docker, two other platforms for virtualization and are trying to utilize Hadoop on their cluster as well.

## IV. THE RPI CLOUD CLUSTER AT THE FUB

The numerous projects involving Raspberry Pis inspired us to set up our own Raspberry Pi Cloud Cluster program at the Free University of Bozen-Bolzano (FUB). We pursue several goals with our project. In the short term we wanted to construct an RPi cluster in an economical and efficient manner. We have succeeded in doing so and report on this work here. The main activities consisted of:

- The design and development of a rack for the cluster that is durable, scalable, and also esthetically pleasing. Two students from the Faculty of Design and Art worked on this task.
- The design, construction, configuration, and testing of a prototype of an RPi cluster consisting of 300 Raspberries. Several students from the Faculty of Computer Science were involved in this task.

Apart from building the cluster we have a number of longer-term goals. First of all, we want to investigate the limitations of an RPi cluster in terms of performance, usability, and maintainability. This will help us in setting up the simulation of a small-scale cloud data center for teaching and research purposes. We also need to find out how well the results we obtain on our cluster can be transferred to the "real world" of cloud computing. Based on the usability study, we also want to identify scenarios in which an RPi cluster can effectively replace traditional computing methods. We are particularly interested in mobility, i.e., how to move clusters to locations where they are needed (e.g. in difficult terrain or in emergency circumstances), and supporting cloud computing in environments that do not have the financial means and/or infrastructure to establish huge data centers (e.g. universities, schools, or developing countries).

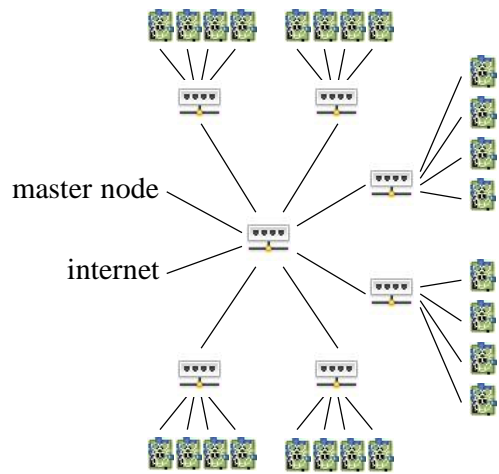

Fig. 2. Network topology

After this high-level overview, in the following sections we cover in more detail the first stage of our project concerning the hardware configuration of the cluster and its system software.

## V. HARDWARE CONFIGURATION

*A. Network Topology*

Our cluster uses a star network topology, with one switch acting as the core of the star and other switches linking the core to the RPIs. Additionally, a master node (more on this later) and an uplink to the internet are connected to the core switch. The system administrators at the Faculty of Computer Science have reserved a /22 network dedicated to the cluster (so we will be able to extend the cluster in the future). Currently, it is hidden behind a firewall and only accessible via a VPN connection. Figure 2 shows a schematic view of the setup. For the other uplinks (the connections between the switches and the core), we plan to use link aggregation for providing at least 2GBit/s of bandwidth for the uplink channels.

*B. Rack*

In an early prototype of our cluster the RPis were taped to the wooden boards of a shelf. As this was not a satisfactory solution, we turned to the Faculty of Design and Art for help, where two students came up with an elegant solution for our problem by designing a rack to hold the Raspberry boards.

This rack has two main components: a holder for mounting individual RPis and a container which can accommodate 24 holders. A holder is made from transparent plastic, which makes it easy to check whether an RPi is switched on and working. See Figure 3 for a photograph: the holder in the background is empty, while the one in the foreground has an RPi fitted to it. Moreover, the holders are independent of each other, making it more convenient to attach and detach individual Raspberries.

Figure 4 shows a completely filled container. These containers are made out of metal and can be stacked on top of each other (note the indentations at the top), leaving space underneath for a switch and a power supply. The capacity of 24 RPis will become evident in a moment, when we discuss the power supply.

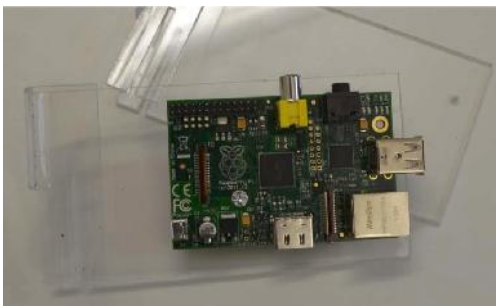

Fig. 3. A holder

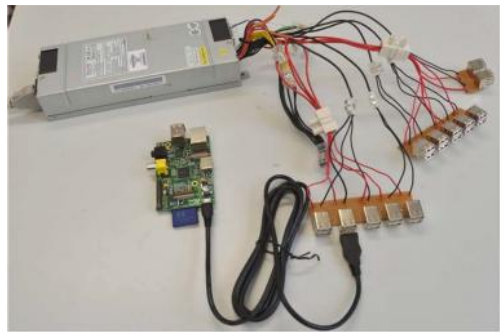

Fig. 5. A modified PSU

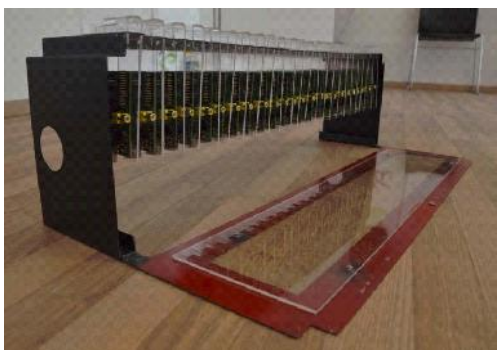

Fig. 4. A container

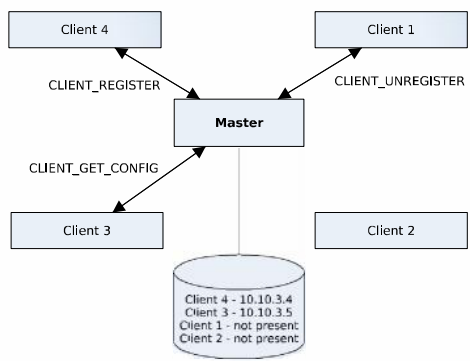

Fig. 6. Registering RPis with the cluster

### C. Power Supply

A standard way of powering an RPi uses a mobile phone charger with a micro-USB interface. While this may be fine for a few boards, it will not scale well to hundreds of RPis: it is quite expensive and we need one power socket per RPi. Having to provide hundreds of power sockets also makes it unwieldy to move racks around. Consequently, we looked around for a better approach and found one by recycling power supply units (PSUs) of broken-down or unused computers. The PSUs we use provide an electric current of 25-30 amp and a Raspberry usually draws a current of 0.5-1.0 amp. To be on the safe side, we connected 24 RPis to each PSU. Figure 5 shows a reconfigured PSU for supplying Raspberry boards with electricity.

## VI. SYSTEM SOFTWARE

### A. Operating System

We created our own OS image based on Debian 7 featuring a minimal configuration sufficient for integrating an RPi into the cluster. This image is copied onto the SD card of a Raspberry, which is then ready to be plugged into the system. Once the node establishes a connection with the master RPi, the master takes care of the full configuration (we describe this in more detail in the next section).

### B. Configuration, Monitoring, and Maintenance

Due to the lack of a Preboot eXecution Environment (PXE) for RPis, we decided to write the software for the low-level configuration, monitoring, and maintenance of the cluster ourselves. In particular, we need to take care of the following tasks. First, we need to keep track of devices joining and leaving the cluster. When a node joins for the first time, it needs to be fully configured: it runs the bootstrap on the SD card, requests the configuration files from the master, and finally registers with the cluster. For fully configured nodes, the master handles requests to register and unregister them. Figure 6 depicts several nodes in different states of registering with the master and being configured. Second, the master RPi also monitors the nodes currently registered with the cluster. Figure 7 shows a snapshot of our monitoring panel. Finally, our tools also allow us to push updates to the individual RPis when upgrading the cluster.

### C. Cloud Services

The set-up described in the previous section is fairly low-level, as it views each RPi as an individual computing unit. Ideally, we would like to run a full-fledged cloud computing stack on our cluster, taking care of managing resources, such as CPUs, storage, and network. While there are freely available open-source cloud computing platforms, such as OpenStack, they are too resource-hungry for an RPi cluster. Consequently, we have to find or come up with our own low-weight solutions for operating a cloud service. The features expected as a cloud platform include centralized management to the resource pool, usage monitoring, automated service provision and online access to the acquired resources. For a start, we have focused on the following aspects of resource management: subcluster organization and storage.

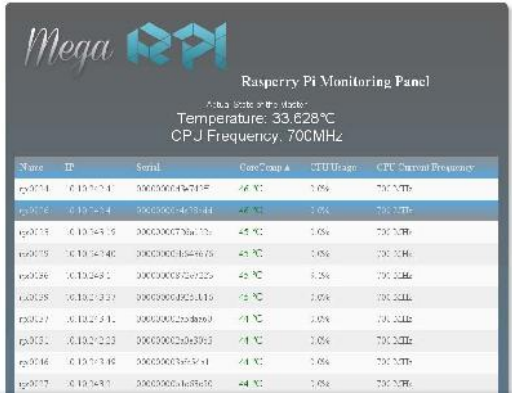

Fig. 7. Monitoring panel

The whole cluster is split into subclusters, allowing different users to run processes in parallel on different parts of the infrastructure. At the moment a subcluster is composed of a predefined set of RPis. It is possible to move an RPi from one subcluster to another one and currently we are working on a solution to be able to make this switch dynamically (i.e., while the system is running) without burdening the system's resources too heavily.

In terms of storage there are two main challenges: the SD cards are slow and the storage they provide is distributed over the whole cluster. Therefore, we decided to use a network storage system to improve the performance of the overall system and to make available a common filesystem for the cluster. We went with a four-bay Network Attached Storage (NAS) from QNAP Systems, which allows us to replace the original firmware with a custom Debian image. This NAS forms part of the master node indicated in Figure 2. Inside the NAS, every subcluster has a dedicated volume managed by LVM (Logical Volume Manager), which is shared by all the RPis belonging to that subcluster. The RPis mount the volume locally via the Network File System (NFS) Version 4. We used NFS rather than iSCSI (Internet Small Computer System Interface) since it allows the sharing of the same volume between different nodes, making inter-node communication via the file system possible. The master node also manages the user authentication in the form of OpenLDAP, granting users access to the nodes of the cluster as well as to the master node. Currently we are not encrypting the general communication inside of the cluster as it is only reachable from the local LAN and security is not one of our top priorities. Once we have finished putting the managing infrastructure into place, we plan to investigate how to add cryptographic protocols to our cluster.

## VII. CONCLUSION AND FUTURE WORK

We have taken the first important steps in getting cloud computing research using live systems off the ground at the Free University of Bozen-Bolzano. A cluster consisting of 300 Raspberry Pis is up and running at the Faculty of Computer Science. In order to reach this point, we had to tackle several challenges, such as supplying power, connecting and housing this large number of RPis, and installing and configuring system software.

For future work we see two paths (which are not necessarily exclusive). On the one hand, we want to put into place light-weight components needed for running cloud services to make our cluster a small-scale, full-fledged, inexpensive, and green testbed for cloud computing research. This includes also understanding its limitations in terms of performance, usability and maintainability. In a first step we plan to benchmark the system in order to determine its overall performance. On the other hand, we plan to experiment with mobile and self-sufficient "cloud in a box" systems by fitting components such as solar panels and satellite uplinks to the cluster. Either way, this will remain a fascinating project, also giving students plenty of opportunities to earn their stripes.

### ACKNOWLEDGMENT

We thank the technicians at the Faculty of Computer Science, Cristiano Cumer, Konrad Hofer, and Amantia Pano, for their support in setting up the Raspberry Pi cluster. We also thank the anonymous reviewers for their helpful and interesting comments and suggestions.